\documentclass[prl,preprint]{revtex4}

\usepackage{epsfig}
\usepackage{dcolumn}
\usepackage{bm}

\begin{document}

\preprint{}

\title{Quantum Fluctuations and the Exchange Bias Field}%

\author{G.J. Mata \email{matag@usb.ve} and E. Pestana}
\affiliation{Departamento de F\'{i}sica, Universidad Sim\'{o}n Bol\'{i}var, Apartado 89000, Caracas 1080A, Venezuela}

\author{Miguel Kiwi}

\affiliation{Facultad de F\'{i}sica, Pontificia Universidad Cat\'{o}lica de Chile,
Casilla 306, Santiago 22, Chile}

\author{Hugues Dreysse}

\affiliation{
IPCMS - GEMME, 23, rue du Loess BP43, 67034 Strasbourg
Cedex 2 France
}

\date{\today}

\begin{abstract}
Ground-state fluctuations reduce the zero-temperature magnetic moments
of the spins in a quantum antiferromagnet. In the neighborhood of surfaces,
interfaces, and other defects which break translational symmetry, these
fluctuations are not uniform. Because of this the magnetic moments of up
and down spins do not exactly compensate each other---as they do in a bulk antiferromagnet.
At a surface or interface this leads to a small magnetic dipole density. The
corresponding dipole field can account for the magnitude of observed
exchange fields. At finite temperatures localized surface (interface) excitations
are populated and change the dipole density. This gives rise to the temperature-dependence
of the exchange field. We expect the fluctuation-induced surface dipole density to
play a role in the magnetic properties of antiferromagnetic nano-particles, as well.
\end{abstract}



\maketitle

Exchange anisotropy arises when two differently ordered magnetic materials,
which are in contact, are cooled in an external magnetic field through their
ordering temperatures. It has been observed, for example, in
clusters or small particles, ferromagnetic (FM) films deposited on
single-crystal or polycrystalline antiferromagnetic (AF) substrates, FM/AF thin
film bilayers, and spin glasses. In these systems the center of the hysteresis loop is shifted
by an amount called the exchange bias
field.  With the convention that the positive field direction is that of the cooling field, this exchange bias
is, in most cases, negative.

Although exchange anisotropy has attracted the attention of physicists and materials
scientists for almost half a century~\cite{meiklejohn:1957,
berkowitz:1999,nogues:1999, kiwi:2001} and has resulted in extensive
technological applications in the storage and sensor industries~\cite{dieny:1991},
a
full understanding of its physical origin has not been
achieved.

From the experimental results, it is now certain
that the effect is due to a fixed spin arrangement
on the antiferromagnetic side of 
the interface~\cite{ohldag:2001,hoffmann:2002,ohldag:2003}.
However, the nature of this arrangement, and the microscopic
mechanism leading to the exhange bias field are still open
questions. Most of the theoretical work has made use of the
classical Heisenberg model in various forms~\cite{kiwi:2001},
an exception being the contribution of Suhl and Schuller~\cite{suhl:1998},
who  have interpreted the exchange field as a self-energy
shift due to the emission and reabsorption of antiferromagnetic spin waves.
Recently, a theory based on the Dzyaloshinsky-Moriya interaction has been
put forward by Ijiry {\it et al}~\cite{ijiri:2004}.
 
In this letter we argue that quantum fluctuations lead to a two-dimensional dipole moment density
at the ferromagnetic-antiferromagnetic interface. The magnetic field due to this dipole density can
explain exchange bias.

Why are quantum fluctuations relevant to the physics of antiferromagnets?
Let us briefly consider this question in the  case of a two-sublattice
antiferromagnet.
The order parameter relevant here is the staggered magnetization:
$
\label{magnetization}
<\mathbf{ \hat{M}}_{AF}>=g\mu_B
< \sum_{\alpha} \mathbf{ \hat{S}}_\alpha -  \sum_{\beta} \mathbf{ \hat{S}}_\beta
>,
$
\noindent where $\mathbf{ \hat{S}}_\alpha $ and $ \mathbf{ \hat{S}}_\beta$ denote the spin
operators at the spin-up and spin-down sub-lattices. If we assume a Heisenberg Hamiltonian
$\mathcal{H} = \sum_{ij} J_{ij}\mathbf{ \hat{S}}_i \cdot \mathbf{ \hat{S}}_j
$, we can see that $[\mathbf{ \hat{M}}_{AF}, \mathcal{H}] \neq 0$, in other
words, that $\mathbf{ \hat{M}}_{AF}$ is not constant in time. Therefore its time-averaged
value must be smaller than its maximum value, even at $T=0$.
(If $N$ is the number of spins and $\sqrt{S(S+1)}$  is the spin magnitude, this
maximum value is  $NS$.)
In a translationally invariant  system the reduction of the magnetization is equally shared
by all spins. Moreover, since the spin up and spin down sublattices are equivalent then
$\sum_{\alpha} \mathbf{ \hat{S}}_\alpha  = -  \sum_{\beta} \mathbf{ \hat{S}}_\beta$
---that is, the system magnetic moment is zero. But when translational symmetry is broken
by surfaces, interfaces, or other defects, a finite magnetic moment can appear because
the decrease of spin-up averages need not equal the corresponding spin-down
reduction~\cite{mata:1985,mata:1990,mata:1994}. In the calculations outlined below
we find that the magnitude of this magnetic moment is only
a few percent of that of a fully uncompensated spin, which is consistent
with the experimental results.

Let us now focus on a specific model, in which atomic
spins of magnitude $\sqrt{S(S+1)}$ are located at the sites of a bcc lattice.
This system is assumed to be divided in two halves by a (001) interface.
To the left of the interface, nearest neighbor spins
are coupled ferromagnetically by
the exchange integral $-J_{F}$. To
the right, nearest neighbor spins are coupled
antiferromagnetically by the exchange integral ${J_{A}}$.
Across the interface, spins
are coupled by the exchange integral
$-J_{0}$. Here we examine three cases, with 
$J_{0}<0$, $J_{0}>0$, and $J_{0}=0$, respectively. The latter corresponds
to a free antiferromagnetic surface.

We decompose the bcc lattice into planes parallel to
the interface. In the antiferromagnetic side, each of these planes
is ferromagnetic and its spin direction alternates from one plane to
the next; we group these planes in pairs an label each {\it pair}
with the index $l\geq0$, in such a way that $l=0$ labels the pair of
planes closest to the interface; for each pair we label the
corresponding planes with the subscript $\alpha $ for spins up, and the label $\beta $
for spins down. In the ferromagnetic side each index ($l<0$) denotes a single, spin up,
layer. This choice of notation reflects the fact that in the
antiferromagnet the unit cell is doubled. The Hamiltonian can be written as:

\begin{eqnarray}
\mathcal{H} & =  & \sum_{l=0}^{+\infty}\sum_{{\bf R},
\delta }
[J_{l,l}{\bf S}_{\alpha}(l, {\bf R}) \cdot {\bf S}_{\beta}(l, {\bf R}+\delta)+
J_{l,l+1}{\bf S}_{\beta}(l, {\bf R}+\delta) \cdot {\bf S}_{\alpha}(l+1,{\bf
R})] + \nonumber\\
 & &  \sum_{l=-\infty}^{-1}\sum_{{\bf R},
\delta }
J_{l,l+1}{\bf S}_{\alpha}(l, {\bf R}+\delta) \cdot {\bf S}_{\alpha}(l+1,{\bf
R}),
\end{eqnarray}

\noindent where
$\mathbf{ R} = a(n_{1}\mathbf{ \hat{x}}+n_{2}\mathbf{ \hat{y}})$
is a two-dimensional lattice point,
$a$ is the lattice constant, $n_{1}$ and $n_{2}$ are
integers,
$\delta = a(\pm {\hat {\bf x}} + \pm {\hat {\bf y}})$,
and
${\bf S}_{a}(l,{\bf R})$ [${\bf S}_{b}(l,{\bf R})$] is a spin in
the
$\alpha $ ($\beta $) plane of the $l$-th pair (and at site ${\bf R}$ in
that
plane).

We now use the Holstein-Primakoff
transformation to rewrite the Hamiltonian in terms of boson
operators $a$ and $b$:

\begin{equation}
S^{z}_{a}(l,{\bf
R})=S-a^{\dag}(l,{\bf R})a(l,{\bf
R}),
\end{equation}

\begin{equation}
S^{+}_{a}(l,{\bf R})={\lbrack
2S-a^{\dag}(l,{\bf R})a(l,{\bf R}) \rbrack}^{1/2}
a(l,{\bf
R}),
\end{equation}

\begin{equation}S^{z}_{b}(l,{\bf
R})=-S+b^{\dag}(l,{\bf R})b(l,{\bf
R}),
\end{equation}

\noindent and

\begin{equation}
S^{-}_{b}(l,{\bf
R})={\lbrack 2S-b^{\dag}(l,{\bf R})b(l,{\bf R}) \rbrack}^{1/2}
b(l,{\bf
R}),
\end{equation}

We neglect spin wave interactions and therefore we
discard quartic and higher order terms. To take advantage of
the translational symmetry we define:
\begin{equation}
a(l,{\bf k})=
{1\over{\sqrt N}}\sum_{\bf R}
a(l,{\bf R}) \exp{({\rm i}{\bf k}\cdot{\bf
R})}
\end{equation}
\noindent and
\begin{equation}
b(l,{\bf k})=
{1\over{\sqrt N}}\sum_{\bf R}
b(l,{\bf R}) \exp{(-{\rm i}{\bf k}\cdot{\bf
R})}.
\end{equation}

The last transformation separates the
Hamiltonian into a sum of
Hamiltonians, one for each value of the
two-dimensional wavector
$\bf k$. Each of these Hamiltonians
can be
thought of as describing a one-dimensional chain, the
parameters of which
depend on $\bf k$ only through the function
$\gamma_{\bf
k}=\cos{(k_{x}/2)}\cos{(k_{y}/2)}$. Once this separation is made,
ground-state properties become sums over $\bf k$. One-dimensional
chains are conveniently analyzed using Green functions, which we now
define:

\begin{equation}
G^{aa}_{ll'}=-{{\rm i}\over \hbar}
\int_{-\infty}^{\infty}
dt \, e^{{\rm i}\omega t} \theta (t)
\langle
[ a(l,{\bf k},t),
a^{\dag}(l',{\bf k},0)]
\rangle
\label{Gf}
\end{equation}

\noindent where $\langle A \rangle$ denotes the
thermal average of $A$ and the operators are in the
Heisenberg representation. We define the functions
$G^{bb}_{ll'}$,
$G^{ab}_{ll'}$, and $G^{ba}_{ll'}$ in the same
fashion.
The average spin at plane $l$ is given
by

\begin{equation}
\langle S^{z}_{l,a}\rangle =S+\sum_{\bf k}{1 \over
\pi}
{\rm Im}\int_{-\infty}^{\infty}d\omega
\frac{G^{aa}_{ll}(\omega,\gamma_{\bf k})}{e^{\hbar \omega / k_{B}T}-1}
\label{deltaS1}
\end{equation}

\begin{equation}
\langle
S^{z}_{l,b}\rangle =-S-\sum_{\bf k}{1 \over \pi}
{\rm Im}
\int_{-\infty}^{\infty}d\omega
\frac{G^{bb}_{ll}(\omega,\gamma_{\bf k})}{e^{\hbar \omega / k_{B}T}-1}
\label{deltaS2}
\end{equation}

The spectral distributions ${{1}\over{\pi}}{\rm Im}G$ in
Equations~\ref{deltaS1}  and~\ref{deltaS2} contain the contributions
of interface and bulk excitations.The interface excitations
are localized in a few layers about interface. They appear
for all values of the model parameters and dominate the spin
reduction, and hence the net magnetization, at the interface.When
$T=0$ the integrands in~\ref{deltaS1}  and~\ref{deltaS2} are zero for
$\omega>0$. The negative $\omega$-axis gives the effect of
virtual spin waves on ground-state fluctuations~\cite{mata:1985}. 
At finite
temperatures real spin waves are excited which further change
the net magnetization.

\begin{figure}
\includegraphics[width=3in,angle=-90]{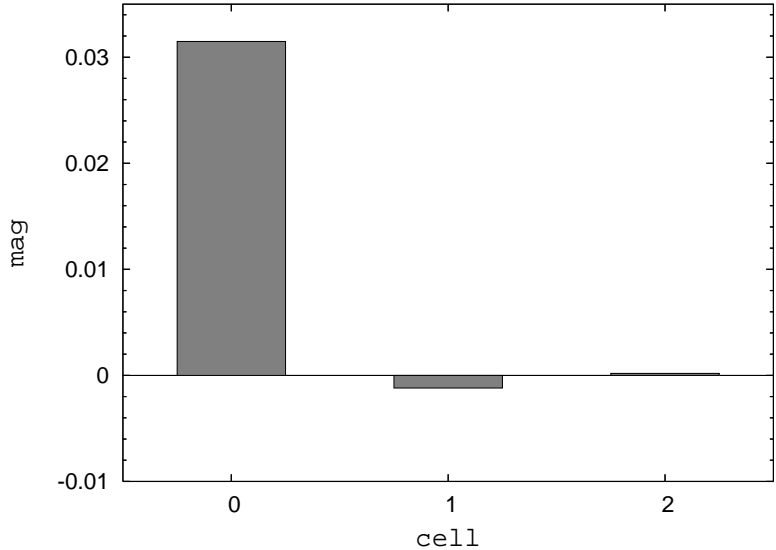}
\caption{Net magnetization per unit cell in the antiferromagnetic side of the interface. The index $cell$
denotes a pair of layers of opposite spins. In units such that
$J_A+J_F=1$, the parameters are $J_A= 0.615$, $J_F= 0.385$, and
$J_0= 0.5$.
}
\label{magnprofile1}
\end{figure}

\begin{figure}
\includegraphics[width=3in,angle=-90]{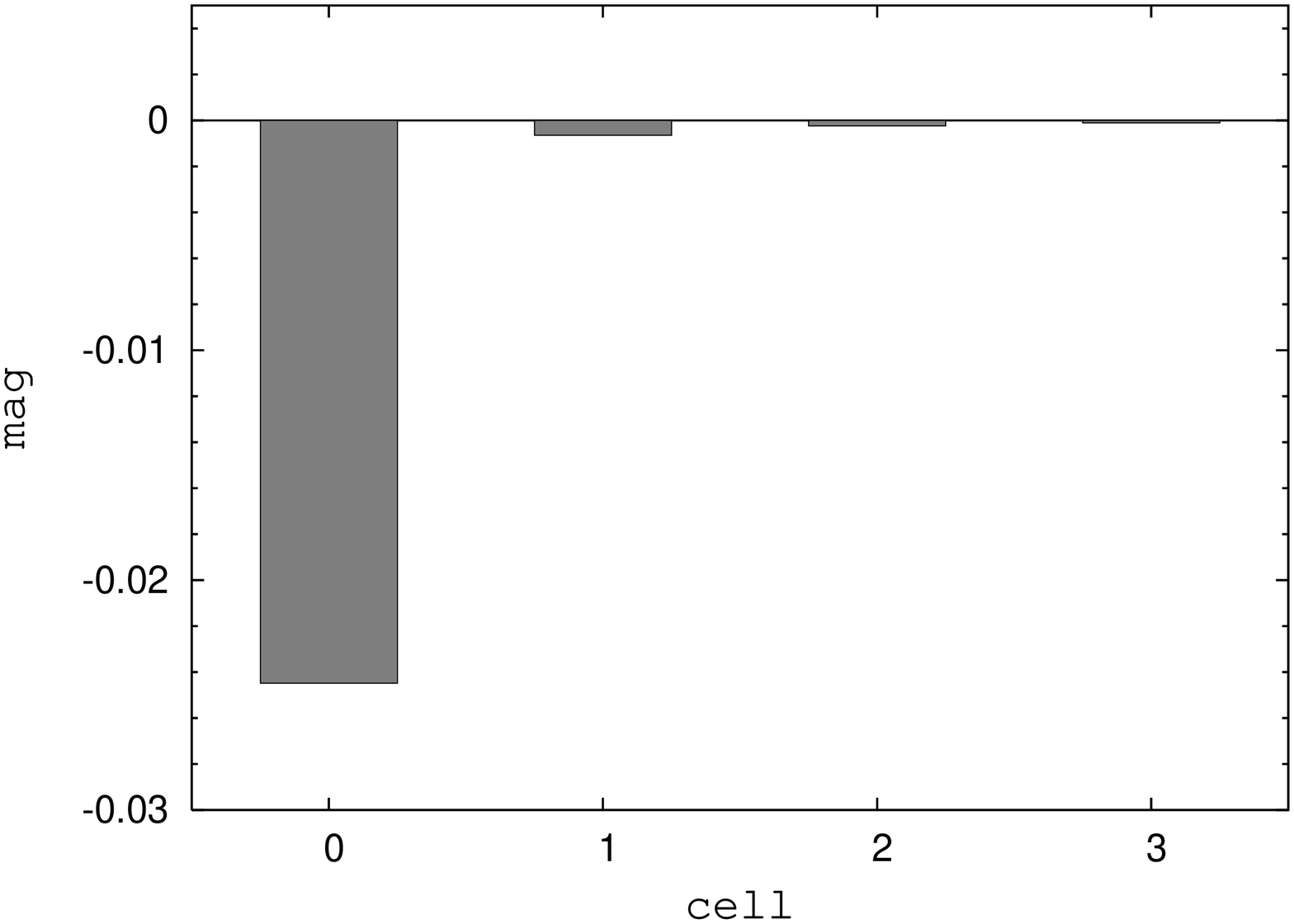}
\caption{Net magnetization per unit cell near the antiferromagnetic side of the interface. The index $cell$
denotes a pair of layers of opposite spins. In units such that
$J_A+J_F=1$, the parameters are $J_{A}= -0.17$, $J_F= 0.83$, and
$J_0= 0.08$.  
}
\label{magnprofile2}
\end{figure}

In figures~\ref{magnprofile1} and \ref{magnprofile2}  we show the magnetization per unit cell in the
antiferromagnetic side of the interface. In both cases, the maximum net magnetization is of 
the order of a few
percent of that of an atomic spin. In figure~\ref{magnprofile1} the coupling across the interface
is ferromagnetic and the net magnetization is parallel to the ferromagnetic moments.
In figure~\ref{magnprofile2} the coupling across the interface
is antiferromagnetic and the net magnetization is antiparallel to that of the ferromagnet. 
In general, we find that for  $| J_0  | \lesssim | J_{A} |$  the net magnetization
of the antiferromagnet is of the order of a few percent of that of an atomic spin. 

The magnetic moments of the antiferromagnetic spins produce
a dipolar magnetic field. This is a long range effect.
We can estimate the magnitude of this field
assuming that the net magnetization is uniformly distributed at a single
layer. If we take this layer to be a disk of radius $R$, the field at a distance
$z$ on the axis is given by $B(z)=(\mu_{0}g\mu_{B}\delta S/Ra^2)
(1+z^2/R^2)^{-3/2}$. With $R\sim 1$ nm, $\delta S\sim 10^{-2}$, and the
lattice constant $a\sim 0.1$ nm, we find that $B(0)\sim10^3$ oersted.

Notice that in figure~\ref{magnprofile1} the exchange bias field is negative,
whereas in figure~\ref{magnprofile2}, it is positive.

\begin{figure}
\includegraphics[width=3in,angle=-90]{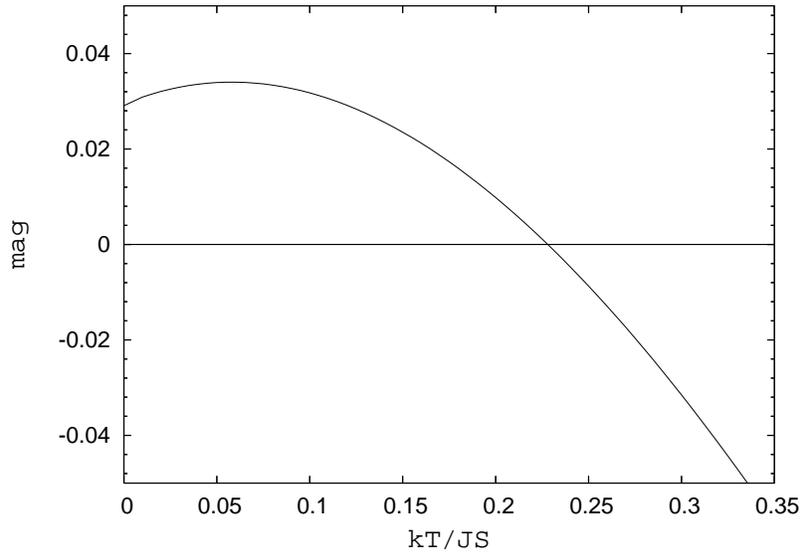}
\caption{Surface magnetization as a function of $k_{B}T/JS$
To model the surface we take $J_0=0$}
\label{magnvsT}
\end{figure}

We now turn our
attention to the free surface case, which we can describe  by simply taking $J_{0}=0$.
In figure~\ref{magnvsT} we show the low temperature variation of the
net surface magnetization. As the temperature rises, surface spin waves
are excited which change the dipole density. 

The magnetic moment is, again, of the order of a few percent of that
of an uncompensated monolayer, as has been observed by Takano
{\it et al.}~\cite{takano:1997} in CoO/MgO multilayers. For this particular system the sign
of the exchange field reverses. Such kind of reversal has been experimentally 
observed~\cite{prados:2002}.

In summary, we have shown that the dipole field generated by uncompensated
quantum fluctuations can account for the observed exchange bias fields. The temperature
dependence is explained in terms of the excitation of surface spin waves which, for low
enough temperatures, can be accounted for by our linear theory. Finally,
since we find that  pure antiferromagnets develop a
surface magnetic moment, our theory could be tested by experimental studies
of clean antiferromagnetic surfaces.

\end{document}